\documentclass[doublecol]{epl2}

\usepackage{amsmath,amssymb}
\usepackage{times}
\newcommand{\grad}{{ \nabla}}

\usepackage[latin1]{inputenc}
\usepackage[english]{babel}

\title{Sedimentation, trapping, and rectification of dilute bacteria}  
  
\author{J. Tailleur \and M. E. Cates}
\institute{SUPA, School of Physics And Astronomy, University of Edinburgh, JCMB
Kings Buildings, Edinburgh EH9 3JZ, United Kingdom}
\date{\today}


\abstract{ The run-and-tumble dynamics of bacteria, as exhibited by
\textit{E. coli}, offers a simple experimental realization of
non-Brownian, yet diffusive, particles. Here we present some analytic
and numerical results for models of the ideal (low-density) limit in
which the particles have no hydrodynamic or other interactions and
hence undergo independent motions. We address three cases:
sedimentation under gravity; confinement by a harmonic external
potential; and rectification by a strip of `funnel gates' which we
model by a zone in which tumble rate depends on swim direction. We
compare our results with recent experimental and simulation literature
and highlight similarities and differences with the diffusive
motion of colloidal particles.}

\begin{document}

\maketitle

Recent years have seen an upsurge of activity at the interface between
physics and biology. In many cases, relatively well-established
physics tools have been used to address problems of major interest to
biologists. In other cases problems that might appear arcane to some
(though not all) biologists have nonetheless posed interesting new
physics questions. For instance, the motion of {\textit{E. coli}} and
similar bacteria involves a series of roughly straight-line `runs'
punctuated by rapid changes of direction
(`tumbles')~\cite{Berg2004}. This run-and-tumble dynamics can, in
idealized form, serve as a paradigm for non-Brownian diffusive motion,
and be used to explore some central concerns of nonequilibrium dynamics,
such as the origins of phase separation in systems without detailed
balance ~\cite{Evansreview,Tailleur2008}.

\textit{E. coli} is a unicellular organism with external flagellar
filaments, which form a bundle at one pole of the
cell~\cite{Berg2004}. When these filaments synchronously rotate
counterclockwise, the organism swims in trajectory that is basically
straight (with small effects of rotational Brownian
motion)~\cite{Berg1973}. A stochastic intracellular event results in a
change of activity \cite{Bray2001,Webre2003} so that one or more
filaments start rotating clockwise; the organism then
tumbles~\cite{Larsen1974, Turner2000}. The tumble is of short
duration; once all filaments start rotating counterclockwise again, a
new run begins.

Supposing the runs to have fixed direction and speed $v$, and the
tumbles to occur instantaneously and randomly at rate $\alpha$ (with
each tumble fully randomizing the direction of motion), it is easily
shown that in free space such run-and-tumble organisms obey the
diffusion equation at large length and time scales, with diffusivity
$D = v^2/\alpha d$ where $d$ is the dimensionality. Yet, because
bacteria are not Brownian particles close to thermal equilibrium, one
cannot write the usual Einstein relations $D= kT \mu$. Here $\mu$ is
the mobility, which controls the mean velocity ${\bf v} = \mu {\bf F}$
of a particle subject to an external force ${\bf F}$; whenever this
force is conservative (${\bf F} = -\nabla U$), the Einstein relation
ensures that, for truly Brownian particles, the equilibrium mean
density $\rho$ obeys the Boltzmann distribution, $\rho \propto
{\exp[-\mu U/D]=\exp[-U/kT]}$.

As will be seen from two of the examples below (sedimentation and the
harmonic trap), forced run-and-tumble particles show more complicated
behaviour than do Brownian ones. Indeed, even when forces are
conservative and a flux-free steady-state density is recovered, this
is not in general a Boltzmann distribution for the potential $U$, even
if $D$ is rescaled at will: there is no general `effective
temperature' concept.

This was discussed briefly in~\cite{Tailleur2008} which however
focussed on the case of \textit{interacting} particles.  In the
present work we explore the noninteracting limit in more detail,
presenting a range of results that include an exact steady-state
calculation of the sedimentation decay length in 3D.  (The
interactions that we omit include hydrodynamics~\cite{goldstein},
which may bring important changes even at modest density.) Oddly, we
have found very little discussion of external force fields in the
previous literature on noninteracting run-and-tumble
particles~\cite{Schnitzer1993,othmer}, much of which assumes
rotational symmetry of the set of possible particle velocities. This
assumption is inspired by models of bacterial chemotaxis, which for
good biological reasons take the tumble rate, and not the swim speed,
to be a function of swim direction. But external forces, such as
gravity, break this symmetry and thus require separate treatment.

As well as external force fields, we consider below the effect of
asymmetric obstacles, such as funnel gates ~\cite{Galajda2007},
arguing that these can be modelled by a zone in which there is
effective asymmetry in the tumbling rates for particles moving in
opposite directions. We show that this viewpoint provides a
semiquantitative account of the rectification experiments of
~\cite{Galajda2007}.

\textit{Effective temperature:} We first consider noninteracting
run-and-tumble particles in 1D with symmetric tumbling rates $\alpha$
and swim speeds $v$. At times long compared to $\alpha^{-1}$ and
length scales large compared to $\ell \equiv v/\alpha$ (the
run-length) the motion is diffusive, with diffusivity
$D=v^2/\alpha$. In the presence of an external potential $U$, the
speeds of left- and right-going bacteria become $v_{L,R}=v\pm \mu
\nabla U$. The diffusion-drift equation for the probability density
$P$ becomes~\cite{Schnitzer1993,Tailleur2008}
\begin{equation}
  \begin{aligned}
    \label{diffapp}
    \partial_t P&=-\nabla {\cdot} J\\ {J}&={-\frac{v^2-\mu^2
	(\grad U)^2}{\alpha} \nabla P - \mu \grad U \left(1- \frac {2\mu\Delta U}
      \alpha\right) P}
  \end{aligned}
\end{equation}
{If the external field only slightly perturbates the swim speed
of the bacterium, that is if $|\mu \grad U| \ll v$, and if it does not
induce large gradients of the velocity field on the scale of the run
length ($|\frac{v \mu \Delta U}\alpha \ll v|$)}, the current becomes
at first order $J=-\frac{v^2}{\alpha} \nabla P - \mu (\nabla U) P$ and
the flux-free steady state then obeys
\begin{equation}
  \label{eqn:efftemp}
  P \propto \exp[-\hat\beta U];\qquad\hat\beta=\frac{\mu \alpha}{v^2}=\frac{\mu}D
\end{equation}
Such 1D bacteria thus behave as ``hot colloids'', with an effective
temperature $1/\hat\beta$; however this breaks down nonperturbatively
\cite{Tailleur2008}. We show below that the same is true for $d>1$.

\textit{Sedimentation:} An obvious physics question concerns the
sedimentation equilibrium of run-and-tumble bacteria: what is the
steady-state probability density $P(z)$ to observe a bacterium of
buoyant mass $m$ at a height $z$ above a hard wall, in the presence of
a downward force $F = -mg$? A technical complication is that the
singular forces arising at the wall have to be described (or replaced
by a rule for modifying nearby trajectories). This will materially
influence the profile in a proximal region whose height is set by the
run length $v/\alpha$. For simplicity we here consider only the distal
part of the profile at much larger $z$.

In 1D the problem can be solved by the methods
of~\cite{Schnitzer1993,Tailleur2008} which developed diffusion-drift
equations for the density that nonetheless exactly recover
flux-free~\cite{Schnitzer1993} or general~\cite{Tailleur2008} steady
states.  A route more readily generalized to $d>1$ instead addresses
the following equations for steady-state probability densities which
contain no approximation beyond those of the defining model:
\begin{eqnarray}
  P(z) dz &\equiv& \sum_{c=\pm 1} P(c,z)dz\label{one}\\ P(c,z)dz &=&
  \frac{\alpha}{2}\int_0^\infty P(z_i(z,c,\tau))dz_i e^{-\alpha\tau}
  d\tau\label{two}
\end{eqnarray}

{Here $P(c,z)$ is the probability density of finding a bacterium
at height $z$ that swims either upwards (labelled $c=1$) or downwards
($c=-1$) and $P(z)$ is the total probability density. Eq.(\ref{two})
then states that a particle of type $c$ currently (at $t=0$, say) in
the small height interval $(z,z+dz)$ last underwent a tumble at some
earlier time. (Note that in 1D, half the tumbles leave $c$ unchanged,
whence the $1/2$ factor). If this event occurred between times $t =
-\tau$ and $t=-(\tau+d\tau)$, the bacterium was then in a height range
$(z_i,z_i+dz_i)$ where $z_i = z-(vc-v_T)\tau$. Here $v_T=-\mu F$ so
that $vc-v_T$ is the net upward velocity of a swimmer of type
$c$. (Note that for constant $F$ only, we also have $dz_i/dz = 1$.)
Thus the contribution to $P(c,z)dz$ of the `age-slice'
($\tau,\tau+d\tau$) is $(\alpha/2)d\tau\times P(z_i)dz_i\times
\exp[-\alpha\tau]$ where the three factors represent respectively the
probability of a tumble event occurring in the relevant time interval;
the probability of finding a bacterium at this time in the required
height range to later arrive at $z$; and the survival probability
against further tumbles at intervening times. Integrating this over
$\tau$ recovers (\ref{two}).}

Summing \eqref{two} over $c$ yields a linear integral equation for
$P(z)$ which can be solved by Fourier transform. Defining $P(\omega)=
\int_{-\infty}^{\infty}P(z)\exp(-i \omega z)dz$, \eqref{one} and
\eqref{two} imply
\begin{equation}
  P(\omega)\, \omega \,[\omega (v^2-v_\tau^2)-i\alpha v_\tau]=0
\end{equation}
\begin{equation}
  \Rightarrow P(\omega)= A \delta(\omega)+B \delta\Big(\omega-\frac {i
  \alpha v_\tau}{v^2-v_\tau^2}\Big)
\end{equation}
Thanks to the flux-free boundary condition at $z=0$, one has $A=0$ and
inverting the Fourier transform yields
\begin{eqnarray}
  P(z)&=& P_0 e^{-\kappa z}\label{three}\\
  \kappa &=& \frac{v_T\alpha}{v^2-v_T^2}\label{four}
\end{eqnarray}
where we have assumed $v>v_T$. In the perturbative regime $v_\tau \ll
v$, one recovers the result \eqref{eqn:efftemp}. More generally,
however, although {Eq.}~\eqref{three} has the Perrin form (a
Boltzmann distribution under gravity) {Eq.}~\eqref{four}
contradicts the result $\kappa = v_T/D$ for Brownian particles. In
particular, the sedimentation length $\kappa^{-1}$ tends to zero as
$v_T$ is raised towards $v$, while for $v_T>v$ both species of
particle have a net downward motion and nothing (short of true
Brownian motion, if present) can save the system from complete
gravitational collapse~\cite{Tailleur2008}.  \if{ The above solution
exactly describes the distal region ($z \gg v/\alpha$) above a
confining wall. The amplitude $P_0$ will then depend on the details of
the wall and on the total density $\intP(z)dz$. }\fi

Unlike the differential equation methods of
\cite{Schnitzer1993,Tailleur2008}, the above integral-equation
approach generalizes, without further approximation, to the 3D
case. By identical reasoning one finds an exact equation for
sedimentation of ideal swimmers
\begin{eqnarray}
P(z) dz &=& 2\pi\int_{-1}^1 dc P(c,z)dz \label{five}\\
P(c,z)dz &=& \frac{\alpha}{4\pi}\int_0^\infty P(z_i(z,c,\tau)dz_i e^{-\alpha\tau} d\tau \label{six}
\end{eqnarray}
where $c =\cos\theta$ is the angle between the propulsion direction
and the vertical, so that the net upward velocity is $vc-v_T$ as
before, and the definition of $z_i(z,c,\tau)$ is likewise
unchanged. Integrating \eqref{two} over $c$ and proceeding as before
via Fourier transform, one gets for $P(\omega)=\int_{-\infty}^\infty
P(z)\exp(i \omega z)$
\begin{equation}
  P(\omega)=P(\omega)\frac \alpha {\omega v} \int_0^\infty
  \exp\Big(\!\!-\frac{\alpha+i\omega v_\tau}{\omega v}\Big) \frac{\sin(u)}u du
\end{equation}
The r.h.s can be computed, for instance by power expansion of the sin
function, to yield
\begin{equation}
  P(\omega) \Big[\frac {\omega v}\alpha- \arctan\Big(\frac {\omega v}{\alpha+i \omega v_\tau}\Big)\Big]=0
\end{equation}
The constant solution $P(\omega)=\delta(\omega)$ is once again
forbidden by the flux-free boundary condition and one is left with
$P(\omega)=\delta(\omega-\omega_0)$ where $ {\omega_0 v}/\alpha=
\arctan({\omega_0 v}/(\alpha+i \omega_0 v_\tau))$. Inverting the
Fourier transform again yields the Perrin form (\ref{three}), where
$\kappa=i \omega_0$ now has to satisfy
\begin{equation}
\ln\left(\frac{\kappa(v_T+v)+\alpha}{\kappa(v_T-v)+\alpha}\right) =
\frac{2\kappa v}{\alpha} \label{seven}
\end{equation}
Just as in 1D, linearization of (\ref{seven}) yields the
quasi-Brownian result ($\kappa = v_T/D$ with $D = v^2/3\alpha$) at
leading order in $v_T$, whereas beyond this, although the Perrin form
(\ref{three}) is maintained, its decay rate $\kappa$ again diverges
when $v_T\to v$ {and is infinite for $v_T>v$}. In this
regime (barring true Brownian motion) gravitational collapse occurs
just as found in 1D, and for the same reason.  Interestingly, since
(\ref{seven}) and (\ref{four}) differ, run-and-tumble diffusion cannot
generally be factorized over orthogonal spatial directions even if, as
here, the forcing is uniaxial and translationally invariant.

\if{Note that the method followed above to compute exact steady-state
in 1 and 3 dimensions could in principle be applied to more general
external forcing, though the corresponding integral equation could
require the use of more advanced methods~\cite{Morse}.}\fi

\begin{figure}
  \centering\includegraphics{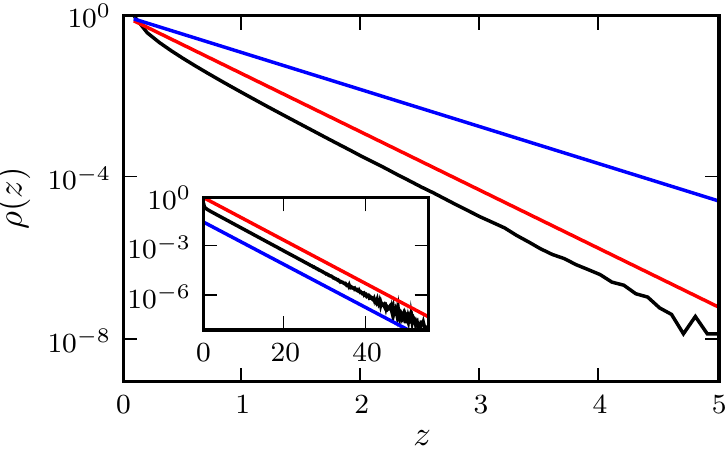}
  \caption{Simulations of 3D run-and-tumble bacteria under
    gravity. Black: full numerics; red: exact solution; blue:
    quasi-Brownian approximation, with vertical offsets introduced for
    clarity. For $v_\tau/v=0.7$, the deviation from the diffusive
    approximation is apparent. {\bf Inset} For $v_\tau/v=0.1$, the
    diffusive approximation works very well.}
  \label{fig:sedim}
\end{figure}

For a typical bacterium in terrestrial gravity one expects
$v_T \simeq 1\mu$ms$^{-1}$ whereas $v\sim
20\mu$ms$^{-1}$~\cite{Berg2004}. This means that the sedimentation
problem (at least in the distal region) remains close to the
quasi-Brownian limit; however the opposite limit could easily be
reached using a centrifuge. We hope that our exact 3D solution will
encourage experiments on bacterial sedimentation. We have also tested our prediction by an explicit stochastic
simulation of an ensemble of run-and-tumble particles, recovering satisfactorily the distal solution (Fig.~\ref{fig:sedim}).

\textit{Trapping:} Another physics question of possible experimental
relevance concerns trapped bacteria. Consider, {\em e.g.}, a harmonic
potential ($U(z) = \lambda z^2/2\mu$) {for} which $v_T(z) =
-\lambda z$. In 1D, (\ref{one},\ref{two}) still hold, but solving
$\dot z = vc - \lambda z$ gives in (\ref{two})
\begin{equation}
\lambda z_i(z,c,\tau) = vc - (vc-\lambda z)\exp[\lambda
\tau]\label{eight}
\end{equation}
so that $dz_i/dz = \exp[\lambda \tau]$. The solution found from the
diffusion-drift formulation ~\cite{Tailleur2008} is, for $|z|<v/k$,
\begin{equation}
P(z) = P(0)\left[1-(\lambda z/v)^2\right]^{\alpha/2\lambda -1}
\label{nine}
\end{equation}
with $P(z) = 0$ for $|z|>v/\lambda$. On substitution in
(\ref{one},\ref{two}) with use of (\ref{eight}), we confirm
(\ref{nine}) to be an exact 1D result.

As noted already in \cite{Tailleur2008} it is not of Boltzmann form
($P\sim \exp[-\alpha \lambda z^2/2v^2]$), nor can it be made so by
any choice of `effective temperature' (any rescaling of $D$). As
discussed earlier, it does however approach the quasi-Boltzmann result
if simultaneously $(k \lambda/v)^2\ll 1$ (defining a region where
$v_T\ll v$) and $\alpha/\lambda \gg 1$ {(i.e. the velocity barely
changes over a run length)}. This additional requirement ensures that
tumbling is frequent on the length scale over which $v_T$ changes
appreciably and that the diffusion-drift approximation \eqref{diffapp}
holds. (Note that our exact result (\ref{nine}) is {\emph{not}}
limited to this case.)  The vanishing of $P$ for $|z| > v/\lambda$
arises, as in sedimentation, because in this region all particles are
moving towards the centre of the trap and a finite density cannot be
maintained in steady state.

\begin{figure}
  \centering\includegraphics{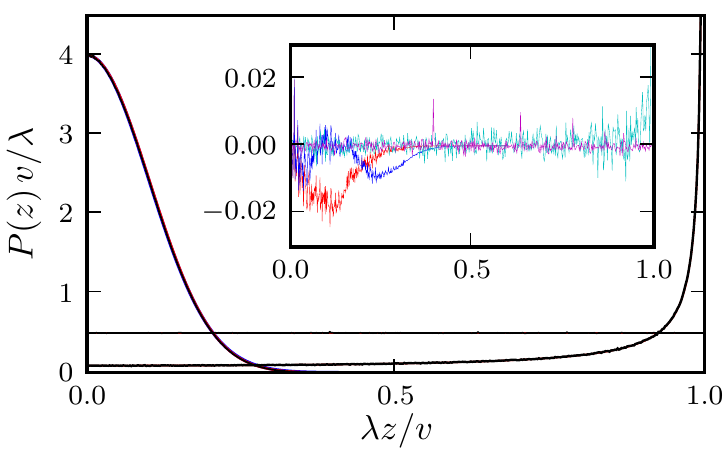}
  \caption{1D run-and-tumble bacteria in a quadratic trap
    ($v=1,\,\alpha=1$). The result of the simulation of run-and-tumble
    dynamics is displayed in the perturbative case ($\lambda=0.01$),
    the flat case ($\lambda=0.5$) and the strong trapping regime
    ($\lambda=5$). The later shows bimodality as predicted. Also
    plotted are the gaussian approximation in blue (for weak trapping)
    and the exact result in red (for all cases). They overlap
    perfectly with the numerics. {\bf Inset} Differences between
    numerics and : in the weak trapping case, the exact
    result (red) and the gaussian approximation (blue); in the flat
    case, the exact result (magenta); in the strong trapping
    case, the exact result (cyan). The agreement is in all cases
    very good, the errors being roughly two order of magnitudes
    smaller that the densities.}
  \label{fig:trap}
\end{figure}

The breakdown of the `effective temperature' concept is much more
severe here than for sedimentation (where the Perrin form was
maintained, albeit with an altered $\kappa$). Whenever
$\alpha/2\lambda$ is not large the distribution is strongly
non-Gaussian; for $\alpha=2\lambda$ it is flat and for $\alpha \leq
2\lambda$ it is bimodal, with an accumulation of bacteria at the outer
edges the trap (Fig.~\ref{fig:trap}). This phenomenology has a simple
intuitive explanation. First, let us note that the motion of right and
left going bacteria amounts to gradient descent in two effective
potentials $U_{L,R}=\pm v z/\mu + \lambda{z^2/(2\mu)}$ whose minima are
$z_{R,L}=\pm v/\lambda$. The average duration of a run is
$\alpha^{-1}$ whereas after reversal of a particle at $z = z_{R,L}$ it
takes time $\ln 2/\lambda \sim \lambda^{-1}$ to reach the centre. For
$\alpha\gg\lambda$, a bacterium tumbles many times before it can cross
the trap and it thus diffuses within the average, quadratic potential
$(U_R+U_L)/2$, giving a Gaussian steady state, $P\sim \exp[-\alpha
kz^2/2v^2]$. On the other hand, when $\alpha\ll \lambda$, bacteria
descend their potential to $z_R$ or $z_L$, spend a long time there and
then tumble. This generates a bimodal distribution whose maxima are at
$z_R$ and $z_L$.

\begin{figure}
  \begin{center}
    \includegraphics{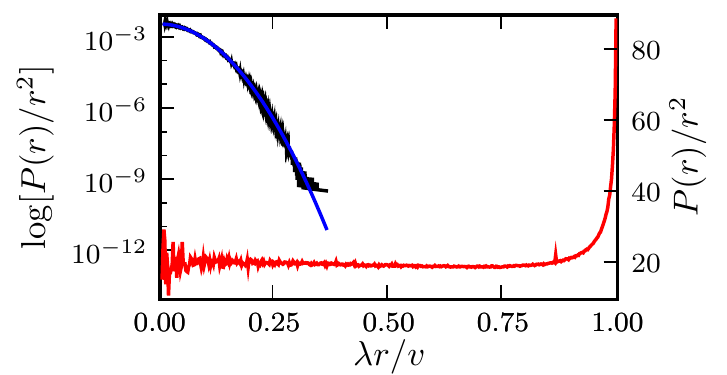}
  \end{center}
  \caption{3D run-and-tumble bacteria in a quadratic trap
    ($v=1,\,\alpha=1$). {\bf Left axis}: weak trapping
    ($\lambda=0.01$).  Numerical results in black coincide with the
    Boltzmann-like approximation in blue $P(r)\simeq r^2 \exp(-3\lambda
    r^2/2)$. {\bf Right axis}: strong trapping case ($\lambda=2$ in
    red), $P(r)/r^2$ increases near the edge of the trap. The bacteria
    accumulate on the surface of the sphere of radius $v/\lambda$.}
  \label{fig:trap3d}
\end{figure}

Stochastic simulations of run-and-tumble particles in 3D indicate a
qualitatively similar behaviour for a radial harmonic trap ${\bf v}_T
= -\lambda\hat{\bf r}$ (Fig.~\ref{fig:trap3d}). Although such a trap
(with trapping radius large compared to a bacterium's diameter) is
hard to realize experimentally ~\cite{CMYW2006}, the same physical
mechanism (with maximal particle density at the outer edge) would
apply for bacteria trapped, say, in an emulsion droplet. It would be
interesting to look experimentally for qualitative deviations from
quasi-Boltzmann statistics in such cases. However, as we address
elsewhere~\cite{rupert} coupling of the particles to a
momentum-conserving solvent could {in some case} drastically
alter, or even destroy, the flux-free steady states predicted here, as
could, on larger time scales, the birth and death of
bacteria~\cite{conrad}.

\begin{figure}
  \begin{center}
    \includegraphics[width=.4\columnwidth]{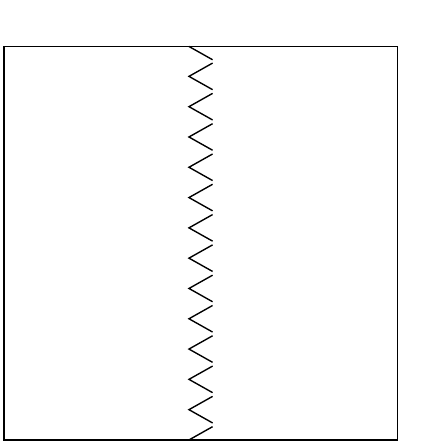}
    \raisebox{.4cm}
	     {\includegraphics{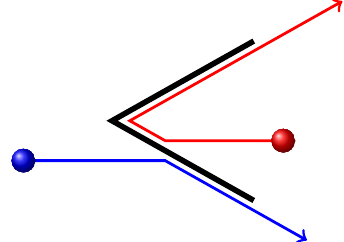}}
  \end{center}
  \caption{{\bf Left:} The sides of the enclosures
    in~\cite{Galajda2007} are 400$\mu m$ long, the arms of the funnels
    are $27 \mu m$ long and their apex angle is $\pi/3$, the funnels
    are separated by gaps $3.8\mu m$ wide. {\bf Right:} After a
    collision with a wall, a particle tends to swim along it. The wall
    thus induces a tumble with fixed outgoing direction.}
  \label{fig:exp1}
\end{figure}

\textit{Rectification:} An intriguing recent experiment has shown how
to trap run-and-tumble bacteria by means quite different from the
external force fields addressed above. This involves `rectification'
of the random swimming motion so that particles move preferentially
into the trap~\cite{Galajda2007}. Two rectangular microfluidic
enclosures contain swimming bacteria (effectively in 2D) separated by
a wall of funnels (See Fig.~\ref{fig:exp1}). Starting from a uniform
density, the system evolves until at long times the two compartments
have uniform populations of bacteria, but with different
densities. The same experiment done with non-swimming bacteria results
in no rectification.

Our model neglects hydrodynamics interactions, both between bacteria
and with the walls~\cite{goldstein}, and simply assumes that a
bacterium swims along a wall after colliding with it, until its next
tumble. (Such motion has been observed in this
experiment~\cite{Galajda2007} and may indeed be of hydrodynamic
origin~\cite{Berg2008}.)  The above fully defines a stochastic model
suitable for simulation (Fig.~\ref{fig:exp1}). To create an
analytical approximation, we represent the wall interactions as an
additional, non-stochastic, reorientation process (`forced tumbling')
whose rate depends on the initial swim direction: while the two
particles in Fig.~\ref{fig:exp1} start symmetrically either side of the
plane of the wall, one will cross this plane whereas the other will
not. A wall of funnels can thus be modelled by a narrow strip in which
(say) left-to-right tumbles are favoured over right-to-left ones.

We next show that this simplified model accounts semi-quantitatively
for the rectification observed in~\cite{Galajda2007}.  First consider
a 1D system with densities $R(x),L(x)$ of right- and left-moving
particles of equal speed $v$ but that tumble with different rates
$\alpha_{R,L}(x)$. Neglecting noise terms~\cite{Tailleur2008} these
obey~\cite{Schnitzer1993,othmer}
\begin{equation}
  \begin{aligned}
    \label{masterequation}
    \partial_t R&= -v\grad R -\frac {\alpha_R}2 R + \frac{\alpha_L}2 L\\
    \partial_t L&= v\grad L -\frac {\alpha_L}2 L + \frac{\alpha_R}2 R
  \end{aligned}
\end{equation}
The $1/2$ factor in front of the tumbling rates formally account for
the fact that after a tumble, bacteria choose their new direction at
random; but any bias in outgoing trajectories can, in 1D, be absorbed
into a redefinition of $\alpha_{R,L}$.

We now divide the system into three regions: for
$x\in[0,L-\epsilon]\cup [L+\epsilon,2L]$, bacteria tumble
symmetrically with a constant rate $\alpha_{R,L}=\alpha_o$, whereas
for $x\in[L-\epsilon,L+\epsilon]$ bacteria swimming to the left are
converted into right-going bacteria with an additional rate
$\alpha_c$, i.e. $\alpha_L=\alpha_o+2\alpha_c$,
$\alpha_R=\alpha_o$. The wall of funnels is thus replaced by a strip
of width $2\epsilon$ in which the \textit{sole effect} is to create a
bias in the tumble rates. Introducing the total probability density
$P=R+L$ and current $J=v(R-L)$, Eqs.\eqref{masterequation} are
equivalent to the continuity equation (\ref{diffapp}) coupled to one
for the current evolution ~\cite{Schnitzer1993,Tailleur2008}
\begin{equation}
  \label{currentdyn}
  \frac 1 \alpha \partial_t J=-D \grad P + V P - J 
\end{equation}
{with $\alpha\equiv(\alpha_R+\alpha_L)/2$; $V\equiv v (\alpha_L -
      \alpha_R)/2\alpha$; and $D \equiv v^2/\alpha$.}
     
In the left and right compartments, where $\alpha_R=\alpha_L$, the
stationary solution of \eqref{diffapp} and \eqref{currentdyn} is
obviously provided by constant profiles, which we denote $\rho_1$ and
$\rho_2$, respectively. In the intermediate strip region however, the
solution is an exponential increase
$\rho(x)\propto \exp\left[x \alpha_c/v\right]$.
Matching these three solutions, one finds for the rectification ratio
\begin{equation}
  \label{eqn:ratio}
  A_{1D}=\frac{\rho_2}{\rho_1}=\exp\left[\frac{
  2\alpha_c\epsilon}v\right]
\end{equation}
Allowing for the inverse dependence of $D$ on dimensionality, a
pseudo-2D version of \eqref{eqn:ratio} would then yield
\begin{equation}
  A_{p2D}=\exp\left[\frac{ 4\alpha_c\epsilon}v\right]
\end{equation}
In fact, a fully 2D analysis is also possible for a specific form of
the tumbling bias. Specifically, if we take constant $v$ and a tumble
rate $\alpha_o + \mbox{\boldmath{$\alpha $}}_1.{\bf u}$ for a swimmer
moving along unit vector ${\bf u}$, then the density $\rho({\bf x})$
obeys in flux-free steady state ~\cite{Schnitzer1993}
\begin{equation}
  \rho({\bf x}) = \rho ({\bf 0})\exp\left[-v^{-1}\int_{\bf 0}^{\bf x}{\bf \alpha_1}({\bf x'}) d{\bf x'}\right] \label{schnitz}
\end{equation} 
Note that single-valuedness of the integral requires $\boldsymbol{ \alpha_1} =
-\boldsymbol\nabla\phi$ for some scalar field $\phi$; without this, there is no
flux-free steady state ~\cite{Schnitzer1993}. For our problem, taking
$\mbox{\boldmath{$\alpha $}}_1$ to be a constant vector normal to the
plane of funnels within the strip and zero outside it, we
{obtain} $A_{2D}=\exp(2 \alpha_1 \epsilon/v)$. Note that the
chosen angular dependence of the tumbling rate does not account
precisely for what happens close to the funnel and we should thus not
expect a perfect quantitative agreement with the experiment.

\begin{figure}
  \centering\includegraphics{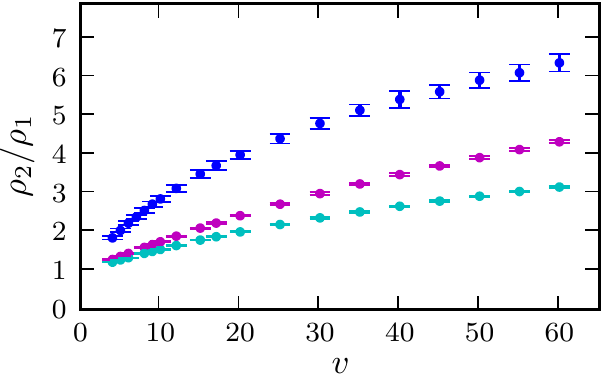}
\vspace{.3cm}

  \centering\includegraphics{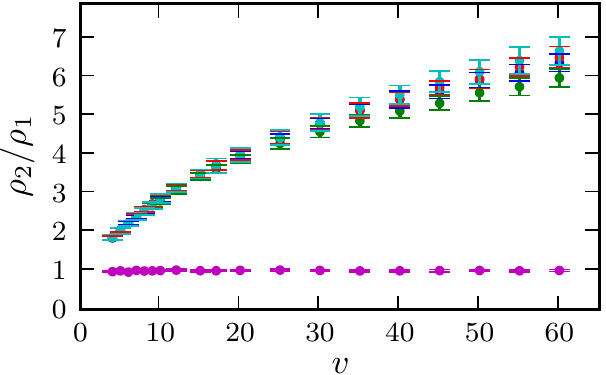}
  \caption{Comparision between theory and simulations (setting
    $\alpha_o=1 s^{-1}$ for the chamber shown in
    Fig.~\ref{fig:exp1}). {\bf Top} The rectification ratio $A$
    measured from numerics (blue, top line), for velocities ranging
    from $4\mu $m s$^{-1}$ to $60 \mu $m s$^{-1}$, is compared with
    $A_{p2d}$ (magenta, middle) and $A_{2d}$ (cyan, bottom). The
    experimental value for the ratio is around $3$~\cite{Galajda2007},
    and is reproduced here for $v\simeq 15\mu $m s$^{-1}$. In the
    experiment, the velocity was measured to be $v=20\pm 5 \mu$m
    s$^{-1}$ and the tumbling rate $\alpha_o=0.75\pm 0.25$
    s$^{-1}$~\cite{GalajdaPrivate}. {\bf Bottom} As predicted by our
    theory, doubling the system width (from $L=400\mu$m in red to
    $L=800\mu$m in blue) does not change the ratio. When doubling the
    height (to $H=800\mu$m in green) and the number of funnels (from
    13 to 26), the curves overlap until the velocity is large enough
    to reveal finite size effects. An asymetric chamber ($L=600\mu$m
    with the right cavity twice as wide as the left one) also yields
    overlapping results (cyan). Last, bacteria colliding elastically
    with the walls are not rectified (magenta).}
  \label{fig:simu1}
\end{figure}

We now compare these theoretical predicitions with our direct
stochastic simulations of 2D run-and-tumble particles in the geometry
described previously. We simulate a square chamber 400$\mu$m wide
{similar to the one used in~\cite{Galajda2007}} (Fig.~\ref{fig:exp1},
with each funnel arm $27 \mu$m long and the gaps $3.8 \mu$m wide. The
apical angle is $\pi/3$. In the simulations, when a swimmer hit a
wall, it aligns with it and proceeds in that direction until the next
tumble. This indeed leads to rectification (Fig.~\ref{fig:simu1}), but
to compare quantitatively with our theoretical model, we must measure
the effective tumbling asymmetry $\alpha_c$ or
$\mbox{\boldmath{$\alpha $}}_1$.

Within the 1D (or pseudo-2D) approach a lower bound for $\alpha_c$ is
provided by the frequency at which particles swimming to the left
encounter the apices of the funnels. Numerically, this can be obtained
as $\alpha_c = n_{col}/t_L^\epsilon$, where one follows a very long
trajectory and measures both $n_{col}$, the total number of such
apical collisions, and $t_L^\epsilon$, the time spent swimming
leftwards within an enclosing strip of size $2\epsilon$. We then
smoothly increased $\epsilon$ until $\epsilon \alpha_c$ converges. In
the fully 2D analysis, integrating the tumble rate
$\alpha_o+\mbox{\boldmath{$\alpha $}}_1.{\bf u}$ over each hemisphere
gives $\langle \alpha_{L,R}\rangle = \alpha_o\pm {2 \alpha_1}/{\pi}$
with only half these tumbles changing the horizontal direction of
swimmers. The corresponding lower bound is then provided by $\alpha_1=
\pi n_{col}/( 2 t_L^\epsilon)$.  The comparisions between
numerics and theory (with these limiting parameter values) are
presented in Fig.~\ref{fig:simu1}. Increasing the swim-speed
increases the flux of bacteria colliding with the funnels, and thus
raises both $n_{col}/t_L^\epsilon$ and the rectification ratio. The
agreement with theory is semiquantitative, and quite satisfactory
given the simplicity of the models. A more detailed analysis of the
angular dependence of the tumbling rate could probably improve the
quantitative agreement, but might not add much further physical
insight.

The `forced-tumbling' picture of rectification admits several
parameter-free predictions. Specifically, the rectification ratio
depends only on the local properties of the funnel wall (gap size,
apical angle, etc.) and not its overall extent or orientation relative
to the walls of the chambers; the size and shape of these chambers are
themselves irrelevant so long as they are large compared to the run
length. (This all follows from the path invariance of the integral in
(\ref{schnitz}).) Some simple checks of this (doubling the chamber
width or height or using an asymetric chamber) are made in
Fig.~\ref{fig:simu1}.

\textit{Detailed balance and symmetry breaking:} Our stochastic
simulations differ significantly from those of Wan et
al.~\cite{Wan2008} who simulated a similar geometry. First, these
authors assumed the tumbling of bacteria to occur at regular rather
than Poissonian time intervals. Our drift-diffusion picture shows this
to be unimportant; this is fortunate since tumbling events for real
bacteria are thought to be near-Poissonian
\cite{Berg1972,Berg2004}. Second, Wan et al. represented each funnel
by a force field deriving from a crescent-shaped potential
barrier. This force field can overcome the swimming force, but cannot
change its direction, whereas the latter is the key element captured
by our own `forced tumbling' description. Omission of this effect
effect may explain why, to achieve the experimental value of $A \simeq
3$~\cite{Galajda2007}, these authors invoked a run-length $v/\alpha_o$
not merely larger than the funnel size but comparable to the size $L$
of their entire chamber. (In this limit, almost every run event
encounters the wall of funnels whereas in the experiment only runs
starting within a strip of width $v/\alpha_o \ll L$ do so.) Our own
simulations instead achieve the observed $A\simeq 3$ with
experimentally plausible parameters ($v\sim 15 \mu $ ms${^{-1}}$,
$\alpha_o =1 $s$^{-1}$).

More fundamentally, it has long been known that to produce directed
motion out of fluctuations (ratchet effect), one needs to break both
space-inversion and time-reversal symmetries~\cite{Ratchet}.  The
spatial symmetry breaking is of course provided by the asymmetric
shape of the funnels themselves; for a wall perforated by symmetric
channels, no rectification can arise~\cite{Galajda2007}. Time reversal
symmetry is more subtle, however. Whereas the motor force of bacteria
does break this symmetry, at the coarse-grained (diffusion-drift)
level, detailed balance is perfectly restored (albeit with respect to
a non-Boltzmann distribution) in both the sedimentation and trap
examples dealt with above. We therefore contend that it is actually
the collisions of bacteria with the funnel walls that break the
time-reversal symmetry in the sense required to create a `ratchet'
from the funnel geometry (Fig.~\ref{fig:exp1}). To check this, we
simulated the same 2D funnel system, with full run-and-tumble
dynamics, in a case where the entrainment of swimmers upon
encountering a wall is replaced by a simple elastic collision
law. Notably, \textit{no rectification whatever} is observed in this
case (Fig.~\ref{fig:simu1}).

Insofar as the rectification problem is subject to Eq.\ref{schnitz},
(see below for exceptions) its steady states are again subject
{to} detailed balance once the diffusion-drift coarse graining is
applied.  The dynamics of approach to states states can be addressed
within the diffusive approximation by neglecting $\dot J$ in
\eqref{currentdyn}. This yields a Fokker-Planck equation whose
late-time relaxation is governed by the first excited state, which for
large systems is a diffusive mode with relaxation time of order of
$\tau=L^2/D$. For a velocity of $15 \mu $m s$^{-1}$ in the geometry of
Galajda et al., this gives a time scale of the order of 10 minutes,
which is quite consistent with the exponential convergence measured
in~\cite{Galajda2007}.

\textit{Steady state fluxes:} The restriction $\mbox{\boldmath{$\alpha
$}}_1({\bf x}) = - \nabla\phi$ encountered in connection with
(\ref{schnitz}) is necessary for steady states in which the flux
vanishes everywhere. It is easy to envisage experiments that violate
this criterion, for instance if walls of uniform funnel density but
both inward and outward orientations are used to create a closed
shape.  If the outward-rectifying sections comprise only a small part
of the perimeter (the exact criterion may depend on shape) the mean
density within the trap will still be higher than outside. Yet, if any
such sections are present, these will carry a steady outward flux,
balanced by an influx elsewhere. Thus the steady state contains
circulating currents. This does not seem to have been looked for
in~\cite{Galajda2007}.

\if{An even simpler example comprises a long toroidal tube containing
one or more funnels walls pointing in the same direction. For obvious
reasons, the run-and-tumble particles will then have a nonzero mean
motion in the direction of rectification. For simplicity we consider
the same chamber geometry as before, but with periodic boundary
conditions in the horizontal direction. We also denote by $x=0$ the
position of the funnel wall, $\ell_F$ its width and $\alpha_c$ the
corresponding enhanced collision rate computed from the case with
closed boundaries. The steady state distribution can be computed by
gluing together a funnel solution (now with nonzero current)
\begin{equation}
  P_F(x\in[0,\ell_F])=\frac{J}{V}+\tilde{A}
  \exp\big[\frac{V x}{D}\big]
\end{equation}
and a linearly decaying profile
\begin{equation}
  P(x\in[\epsilon,2L])=P_1-\frac{ J}{D}x
\end{equation}
where $J$ is the current (which is uniform in space). There are three
unknowns, $J,P_1$ and $\tilde{A}$, which can be found from
normalization and continuity at $x=0,\epsilon$. Using the pseudo-2D
model introduced earlier, one gets for the current
\begin{equation}
  \label{eqn:current2}
  J=2\left[\ell_F \frac{4 L - \ell_F}{v \log
  R}+ \frac{(2L-\ell_F)^2\alpha_o}{v^2 (1-\frac 1 R)}-\frac {\alpha_o} {v^2}
  \frac{(2L-\ell_F)^2}{2}\right]^{-1}
\end{equation}
A precise value for $\alpha_c$ can be obtained from the rectification
ratio of the densities in the flux-free cases. Injecting this in
\eqref{eqn:current2} and comparing with the current $J_m$ actually
measured in simulation of the periodic boundary case allows to check
for consistency of the model. The result is $J/J_m = ****$
WHAT. Agreement is again semiquantitative; but the deviation from
unity shows that our forced tumbling picture (with flux-independent
$\alpha_c$) does not capture all aspects of the rectification
problem.}\fi

\textit{Conclusion:} We have calculated and discussed in this letter
several steady-state density distributions within an idealized model
\cite{Schnitzer1993} of noninteracting run-and-tumble bacteria; our
results included a new exact solution for 3D sedimentation. For
particles in harmonic traps the mechanism causing extremely
non-gaussian density profiles at large ratios of run-length to trap
size was elucidated.  The phenomenon of rectification of
run-and-tumble bacteria by funnel walls was addressed using an
asymmetric tumble rate within both a 1D and 2D model, giving
semiquantitative agreement with direct stochastic simulations and,
using plausible parameters, with the experiments of
~\cite{Galajda2007}. We found no rectification in the case where each
particle collides elastically with a funnel wall, rather than swimming
parallel to the wall until its next tumble. This suggests, at odds
with ~\cite{Wan2008}, that wall interactions (rather than the
intrinsic asymmetry of bacterial propulsion mechanisms) are the
fundamental source of time-reversal asymmetry at the coarse-grained
level relevant to the rectification experiments of~\cite{Galajda2007}.

{\bf Acknowledgments:} We thank R. Austin, R. Blythe, P. Galajda,
J. Kurchan, J.-F. Joanny, S. Ramaswamy and R. Nash for discussions and
EPSRC (EP/E030173 and GR/T11753) for funding. MEC holds a Royal
Society Research Professorship.


\begin{thebibliography}{99}

\bibitem{Berg2004} H. C. Berg, {\em E. coli in Motion}, Springer, NY
  (2004)

\bibitem{Berg1972} H.C. Berg and D.A. Browm, Nature {\bf 239}, 500
  (1972)

\bibitem{Tailleur2008} J. Tailleur, M. E. Cates, Phys. Rev. Lett. {\bf
100}, 218103 (2008)

\bibitem{Evansreview} R. A. Blythe, M. R. Evans, J. Phys. A {\bf 40},
 R333-R441 (2007)

\bibitem{Berg1973} H.C. Berg and R.A. Anderson, Nature {\bf 245}
  380-382 (1973)

\bibitem{Bray2001} D. Bray, \textit{Cell movements}, 2nd edn. Garland,
NY (2001)

\bibitem{Webre2003} D. Webre, P. Wolanin and J. Stock,
Curr. Biol. {\bf 13}, R47-R49 (2003)

\bibitem{Larsen1974} S.H.R. Larsen, et al., Nature {\bf 249}, 74-77
(1974)

\bibitem{Turner2000} L. Turner, W.S. Ryu and H.C. Berg,
J. Bacteriol. {\bf 182}, 2793-2801 (2000)

\bibitem{goldstein} A. Sokolov, I. S. Aranson, J. O. Kessler and
  R. E. Goldstein, Phys. Rev. Lett. {\bf 98}, 158102 (2007);
  C. Dombrowski, et al., Phys. Rev. Lett. {\bf 93}, 098103 (2004);
  T. Ishikawa and T. J. Pedley, Phys. Rev. Lett. {\bf 100}, 088103
  (2008)

\bibitem{Schnitzer1993} M. J. Schnitzer, Phys. Rev. E {\bf 48},
2553-2568 (1993)
    
\bibitem{othmer} T. Hillen and H. G. Othmer, SIAM J. Appl. Math.  {\bf
  61}, 751-775 (2000); H. G. Othmer and T. Hillen, SIAM
  J. Appl. Math. {\bf 62}, 1222-1250 (2002); R. Erban and
  H. G. Othmer, SIAM J. Appl. Math. {\bf 65}, 361-391 (2005)

\bibitem{Galajda2007} P. Galajda, J. Keymer, P. Chaikin, R. Austin,
J. Bacteriol. {\bf 189}, p. 8704-8707 (2007); P. Galajda,et al., J. Modern Optics {\bf 55}, 3413-3422
(2008)

\bibitem{CMYW2006} S. Chattopadhyay, R. Moldovan, C. Yeung, X.L. Wu,
Proc. Natl. Acad. Sci. USA {\bf 103}, 13712 (2006)

\bibitem{rupert} R. Nash, et al., work in progress.

\bibitem{conrad} C. Barrett-Freeman, M. R. Evans, D. Marenduzzo, and
W. C. Poon, Phys. Rev. Lett. {\bf 101}, 100602 (2008)
    
\bibitem{Berg2008} A.P. Berke, L. Turner, H.C. Berg, E. Lauga,
Phys. Rev. Lett. {\bf 101}, 038102 (2008)

\bibitem{GalajdaPrivate} {P. Galajda, private communication}

\bibitem{Wan2008} M.B. Wan, C.J. Olson Reichhardt, Z. Nussinov,
C. Reichhardt, Phys. Rev. Lett. {\bf 101}, 018102 (2008)
    
\bibitem{Ratchet}
%
M.O. Magnasco, Phys. Rev. Lett. 71, 1477 (1993)
%
J. Prost, J.-F. Chauwin, L. Peliti and A. Ajdari, Phys. Rev. Lett. 72,2652 (1994);
%

\end{thebibliography}
\end{document}